\documentclass{article}
\usepackage{epsfig}
\newcommand{\bobfig}[1]{
\mbox{\raisebox{-4mm}{\epsfig{file=#1.EPS,height=9mm}}}}
\newcommand{\bookfig}[5]{
\begin{figure}\centering\fbox{\epsfig{file=#1.EPS,height=#5cm}}
\caption[#2]{#4}\label{#3}\end{figure}}
\def\semi{\mathop{>\!\!\!\triangleleft}}

\begin{document}
\title{Lessons from Quantum Field Theory\\
{\normalsize Hopf Algebras and
Spacetime Geometries}}
\author{A.~{Connes}\thanks{connes@ihes.fr}
\and D.~{Kreimer}\thanks{kreimer@ihes.fr}}
\date{IHES/M/99/22, hep-th/9904044}
\maketitle
\begin{center}
{\em\small Institut des Hautes Etudes Scientifique\\} {\small
Le Bois-Marie, 35 route de Chartres\\ 
F-91440 Bures-sur-Yvette, France\\[5mm]}
 {\sl  We dedicate this paper to Mosh\'e
Flato.}
\end{center}
\begin{abstract}
We discuss the prominence of Hopf algebras in recent progress in
Quantum Field Theory. In particular, we will consider the Hopf
algebra of renormalization, whose antipode turned out to be the
key to a conceptual understanding of the subtraction procedure. We
shall then describe several occurences of this, or closely related
Hopf algebras, in other mathematical domains, such as foliations,
Runge-Kutta methods, iterated integrals and multiple zeta values.
We emphasize the unifying role which the Butcher group, discovered
in the study of numerical integration of ordinary differential
equations, plays in QFT.
\end{abstract}
{\bf \small keywords:} {\small Quantum Field Theory,
Noncommutative Geometry, Renormalization, Hopf Algebras,
Foliations, ODE.\\} {\bf \small MSC91} {\small
81T15,16W30,46L87,58B30,58F18,34Axx.}
\section{Introduction}
The {\em Leitmotiv} of this survey paper is our belief that in some
way the true geometry of spacetime is actually dictated by
quantum field theories as currently used by particle physicists in
the calculation of radiative corrections.

There are two major ingredients in this use of the theory, the
first is the renormalization technique, with all its combinatorial
intricacies, which is perfectly justified by its concrete physical
roots and the resulting predictive power. The second is the
specific Lagrangian of the theory, the result of a long dialogue
between theory and experiment, which, of course, is essential in
producing meaningful physical results.

This Lagrangian is a unique mixture of several pieces, some very
geometrical and governed by the external symmetry group of the
equi\-va\-lence principle, i.e.~the diffeomorphism group ${\bf
Diff}$, the others governed by the internal group of gauge
transformations ${\bf Gauge}$. The overall symmetry group is the
semi-direct product ${\bf G}={\bf Gauge} \semi{\bf Diff}$, which
is summarized by the following exact sequence of groups
\[
1\to{\bf Gauge}\to{\bf G}\to {\bf Diff}\to 1.
\]

But there is more group structure than the external and internal
symmetries of this Lagrangian of gravity coupled with matter. Our
goal in this paper is to explain how the understanding of the
group-theoretic principle underlying the working machine of
renormalization \cite{dhopf} should allow one to improve the
understanding of the geometrical nature of particle physics which
was proposed in \cite{alain1}.

The main point in the latter proposal is that the natural symmetry
group ${\bf G}$ of the Lagrangian is isomorphic to the group of
diffeomorphisms ${\bf Diff}(X)$ of a space $X$, provided one
stretches one's geometrical notions to allow slightly
noncommutative  spaces. Indeed, the automorphism group ${\bf
Aut}({\cal A})$ of a noncommutative algebra ${\cal A}$ which
replaces the diffeomorphism group of any ordinary manifold  has
exactly this very feature of being composed of two pieces, one
internal and one external, which, again, can be given equivalently as
an exact sequence of groups
\[ 1\to{\bf Int}({\cal
A})\to{\bf Aut}({\cal A})\to {\bf Out}({\cal A})\to 1.
\]
In the general framework of NCG the confluence of the two notions
of  metric and fundamental class for a manifold led very naturally
to the equality
\begin{equation}
ds=1/D,\label{E1}
\end{equation}
which expresses the line element $ds$ as the inverse of the Dirac
operator $D$, hence under suitable boundary conditions as a
propagator. The significance of $D$ is two-fold. On the one hand,
it defines the metric by the above equation, on the other hand its
homotopy class represents the K-homology fundamental class of the
space under consideration. While this new geometrical framework
was immediately useful in various mathematical examples of
noncommutative spaces, including the noncommutative torus
\cite{alain2,alain3}, it is obvious that it cannot be a
satisfactory answer for spacetime precisely because QFT will
unavoidably dress the free propagator, as Fig.(\ref{dress})
indicates. It nevertheless emphasizes that spacetime itself ought
to be regarded as a derived concept, whose structure in we believe
follows from the properties of QFT if one succeeds describing the
latter in purely combinatorial terms.
\bookfig{}{!}{dress}{The dressed line element is obtained
from the transition of the bare to the dressed propagator.}{0.5}

Whereas it was simple in the undressed case to recover the
standard ingredients of intrinsic geometry directly from the Dirac
propagator and the algebra ${\cal A}$, it is much more challenging
in the dressed case, and we shall naturally propose the 
Schwin\-ger-Dy\-son equation for the full fermion propagator as the proper
starting point in the general case.

Let us now describe various appearances of Hopf algebras relevant
to Quantum Field Theory. We will start with the Hopf algebra of
renormalization, whose antipode turned out to be the key to a
conceptual understanding of the subtraction procedure. We shall
then describe several occurences of this, or closely related Hopf
algebras, in other mathematical domains, such as foliations
\cite{CM}, Runge-Kutta methods \cite{Br}, iterated integrals
\cite{chen,dchen} and multiple zeta values \cite{sasha}.

We will finally address the Schwinger Dyson equation as a
prototype of a generalized form of an ordinary differential
equation (ODE).
\section{The pertinence of Hopf algebras}
\subsection{Renormalization and the antipode in the algebra of rooted trees}
Renormalization occurs in evaluating physical observable
quantities which in simple terms can be written as formal
functional integrals of the form
\[
\int e^{-{\cal L}(\phi,\partial \phi)}P(\phi,\partial\phi)[d\phi],\;\;\;
{\cal L}={\cal L}_0+{\cal L}_I.
\]
Computing such an integral in perturbative terms leads to a formal
power series, each term $G$ of which is obtained by integrating a
polynomial under a Gaussian $e^{-{\cal L}_0}$. Such Feynman
diagrams $G$ are given by multiple integrals over
spacetime or, upon Fourier transformation, over momentum space,
and are typically divergent in either case.
The renormalization
technique consists in adding counter\-terms ${\cal L}_G$ to the
original Lagrangian ${\cal L}$, one for each diagram $G$, whose
role is to cancel the undesired divergences. In good situations to
which we are allowed to restrict ourselves if we take guidance
from nature, this can be done by local counterterms, polynomials
in fields and their derivatives.

The main calculational complication of this subtraction procedure
occurs for diagrams which possess non-trivial subdivergences,
i.e.~subdiagrams which are already divergent. In that situation
the procedure becomes very involved since it is no longer a simple
subtraction, and this for two obvious reasons: i) the divergences
are no longer given by local terms, and ii) the previous
corrections (those for the subdivergences) have to be taken into
account.

To have an example for the combinatorial burden imposed by these
difficulties  consider the problem below of the
renormalization of a two-loop four-point function in massless
scalar $\phi^4$ theory in four dimensions, given by the following
Feynman graph:
\[
\Gamma^{[2]}  =  \bobfig{}.
\]
It contains a divergent subgraph
\[
\Gamma^{[1]}=\bobfig{}.
\]
We work in the Euclidean framework and introduce a cut-off
$\lambda$ which we assume to be always much bigger than the square
of any external momentum $p_i$. Analytic expressions for these
Feynman graphs are obtained by utilizing a map $\Gamma_\lambda$
which assigns integrals to them according to the Feynman rules and
employs the cut-off $\lambda$ to the momentum integrations. Then
$\Gamma_\lambda[\Gamma^{[1,2]}]$ are  given by
\[
\Gamma_\lambda[\Gamma^{[1]}](p_i)=\int d^4k
\frac{\Theta(\lambda^2-k^2)}{k^2}\frac{1}{(k+p_1+p_2)^2},
\]
and \[ \Gamma_\lambda[\Gamma^{[2]}](p_i)=\int d^4l
\frac{\Theta(\lambda^2-l^2)}{l^2(l+p_1+p_2)^2}
\Gamma_\lambda[\Gamma^{[1]}(p_1,l,p_2,l)].
\]
It is easy to see that $\Gamma_\lambda[\Gamma^{[1]}]$
decomposes into the form $b
\log\lambda$ (where $b$ is a real number)
plus terms which remain finite for
 $\lambda\to\infty$, and hence will produce a
divergence which is a non-local function of external momenta
\[
\sim\log\lambda\int d^4l
\frac{\Theta(\lambda^2-l^2)}{l^2(l+p_1+p_2)^2}\sim\log\lambda\;\log(p_1+p_2)^2.
\]
Fortunately, the counterterm ${\cal L}_{\Gamma^{[1]}}$
$\sim\log\lambda$ generated to subtract the divergence in
$\Gamma_\lambda[\Gamma^{[1]}]$ will precisely cancel this
non-local divergence in $\Gamma^{[2]}$.

That these two diseases actually cure each other in general is a
very non-trivial fact that took decades to prove \cite{BPHZ}. For
a mathematician the intricacies of the proof and the lack of any
obvious mathematical structure underlying it make it totally
inaccessible, in spite of the existence of a satisfactory formal
approach to the problem \cite{EG}.

The situation is drastically changed by the understanding of the
mechanism behind the actual computations of radiative corrections
as the antipode in a very specific Hopf algebra, the Hopf algebra
${\cal H}_R$ of decorated rooted trees discovered in
\cite{dhopf,CK,doverl}.

This algebra is the algebra of coordinates in an affine nilpotent
group $G_R$, and all the non-linear aspects of the subtraction
procedure are subsumed by the action of the
antipode, i.e.~the operation $g\to
g^{-1}$ in the group $G_R$.

This is not only very satisfactory from the conceptual point of
view, but it does also allow the automation of 
the subtraction procedure
to arbitrary loop order (see \cite{BK} for an example of Feynman
graphs with up to twelve loops) and to describe the more elaborate
notions of renormalization theory, -change of scales,
renormalization group flow and operator product expansions-, from
this group structure \cite{BK,dchen,DelKr}.

In \cite{CK} we characterized this Hopf algebra of
renormalization, ${\cal H}_R$, abstractly as the universal
solution for one-dimensional Hochschild cohomology. The Hopf
algebra ${\cal H}_R$ together with the collecting map
(cf.\cite{CK}) $B_+:{\cal H}_R\to {\cal H}_R$ is the universal
solution of the following Hochschild cohomology equation. We let
${\cal H}$ be a commutative (not necessarily cocommutative) Hopf
algebra together with a linear map $L:{\cal H}\to {\cal H}$. The
Hochschild equation is then $bL=0$, i.e.
\[
\Delta L(a)=L(a)\otimes 1+(id\otimes L)\Delta(a).
\]
We shall see later the intimate relation between this problem and
the generalized form of ODE's.

Renormalization can now be summarized succinctly
by saying that it
maps an element $g\in G_R$ to another element $g_o^{-1}g\in G_R$.
Typically, $g$ is associated with bare diagrams and $g_o$ is a
group element in accordance with a chosen renormalization scheme.
In the ratio $g_o^{-1}g$ the undesired divergences drop out.
Hence, $g_o^{-1}$ delivers the counterterm and $g_o^{-1}g$ the
renormalized Green function. It is precisely this  group law which
allowed the description of the change of scales and renormalization
schemes in a comprehensive manner \cite{dchen}. Indeed, if
$g_o^{-1}g$ is one renormalized Green function, and
$g_o^{-1}g^\prime=(g^{-1}_og)(g^{-1}g^\prime)$ another, then we
immediately see the group law underlying the evolution of the
renormalization group.
\subsection{Foliations}
Quite independently, the problem of computation of certain
characteristic clas\-ses coming from differential operators on
foliations led \cite{CM} to a very specific Hopf algebra ${\cal
H}(n)$ associated to a given integer $n$, which is the codimension
of the foliation.

The construction of this Hopf algebra was actually dictated by the
commutation relations of the algebra ${\cal A}$ (corresponding to
the noncommutative leaf space) with the operator $D$
(cf.(\ref{E1})) describing the transverse geometry in the sense of
(\ref{E1}). The structure of this Hopf algebra revealed that it
could be obtained by a general procedure from the  pair
of subgroups $G_0,G_1$ of the group ${\bf Diff}({\bf R}^n)$ of
diffeomorphisms of ${\bf R}^n$: One lets $G_0$ be the subgroup of
affine diffeomorphisms, while $G_1$ is the subgroup of those
diffeomorphisms which fix the origin and are tangent to the
identity map at this point. This specific structure of a Hopf
algebra not only led in this case to a solution of the
computational problem, but also provided the proper generalization
of Lie algebra cohomology to the general Hopf framework in the
guise of cyclic cohomology (see the paper {\em Cyclic cohomology
and Hopf algebras} by Connes and Moscovici in this volume).

In \cite{CK} we established a relation between the two Hopf algebras
which led us to extend various results of \cite{CM}
to the Lie algebra ${\cal L}^1$ of rooted trees, in particular we
extended ${\cal L}^1$ by an
 additional generator $Z_{-1}$, intimately related to natural
growth of rooted trees.

Let us mention at this point that the extended Lie algebra ${\cal
L}$ has a radical as well as a simple quotient. The former is
nilpotent and infinitely generated, the latter turns out to be
isomorphic to the Lie algebra of ${\bf Diff}({\bf R})$, thus
exhibiting an intrinsic relation between the two groups.

Extension of this relation between ${\cal H}(1)$ and ${\cal H}_R$
leads to the Hopf algebra of decorated planar rooted trees,
explored by R.~Wulkenhaar \cite{W}, whose new feature is that the
decoration of the root provides the information necessary to
define the notion of parallel transport, while the decorations of
other vertices are mere spacetime indices.
\subsection{Numerical Integration Methods}
Our description in \cite{CK} of the simplest realization of the
Hopf algebra ${\cal H}_R$ and its related group, together with the
elaboration of this group structure in \cite{BK,dchen}, allowed
C.~Brouder \cite{Br} to recognize a far reaching relation with the
combinatorics of numerical analysis  as worked out by Butcher
several decades ago.

Indeed, Butcher, in his seminal work on the algebraic aspects of
Runge-Kutta methods (and other numerical integration methods)
\cite{Bu} discovered the very same Hopf algebra and group $G_R$.

Let us first remind the reader of the simplest numerical
integration methods in the integration of a given ordinary
differential equation
\begin{equation}
\dot{y}=f(y),\;y(0)=y_0,
\end{equation}
where $t\to y(t)$ is a curve in Euclidean space $E$. The Runge
method, designed to approximate the value $y(h)$ of $y$ at $t=h$
is given by the formula,
\begin{equation}
\hat{y}(h)=y_0+h f(y_0+\frac{h}{2}f(y_0)).
\end{equation}
The virtue of this simple iteration of $f$ is that the Taylor
expansion of $\hat{y}$ for small $h$ agrees up to order two with
the Taylor expansion of the actual solution $y(h)$.

More generally a Runge-Kutta method is an iterative procedure,
governed by two sets of data, traditionally denoted by $a$ and $b$, where
$b_i$, $i=1,\ldots,n$, and $a_{ij}$ are scalars while the implicit
equations are
\begin{eqnarray}
X_i(h) & = & y_0+h\sum a_{ij} f(X_j(h)),\\ \hat{y} & = & y_0+h
\sum b_i f(X_i(s)).
\end{eqnarray}
Their solution is known to be a sum over rooted trees, involving
the very combinatorial quantities ascribed to rooted trees by
the study of QFT, namely tree factorials, CM weights and symmetry
factors \cite{Bu,Br,BK,dchen}.
It has been known since Cayley \cite{Cayley} that rooted
trees are the correct way to label polynomials in higher
derivatives of a smooth map $f:E\to E$. For instance an $n$th
derivative $f^{(n)}$ corresponds to a vertex with $n$ adjacent
branches, as in Fig.(\ref{f1}). \bookfig{}{!}{f1}{The tree
structure of derivatives.}{2}

What Butcher discovered is that Runge-Kutta methods naturally form
a group, whose elements are actually scalar functions of rooted
trees. He gave explicit formulas for the composition of two
methods as well as for the inverse method. He also showed how the
standard solution of a differential equation is obtained from a
particular (continuous) method which he called the Picard method.
There is a nuance between Runge-Kutta methods and the Butcher
group of scalar functions of rooted trees, but Butcher proved that
the Runge-Kutta methods are dense in the latter group.

In \cite{dchen} scalar functions of rooted trees were used to
parametrize and generalize iterated integrals, allowing a
unified description of renormalization schemes. This immediately
allowed C.~Brouder to identify the above group $G_R$ (in the
simplest undecorated case) with the Butcher group.

The supplementary freedom in constructing group elements or 
operations on them provided by the Runge-Kutta description matches
the freedom to choose a renormalization scheme or to describe the
change of a chosen  scheme. Hence the group product, as well as
the counterterm, -the inverse group element, have immediate and
equally elegant counterparts in the Runge-Kutta language worked
out recently by Brouder \cite{Br}.

Moreover, comparing the data $a,b$ of a Runge-Kutta method with our
characterization of the Hopf algebra ${\cal H}_R$ by the
Hochschild cohomology problem leads us to the following natural
framework for the formulation of a universal differential equation
(which also covers the case of the Schwinger-Dyson equation)
given a non-linear map $f:E\to E$. The only nuance between our
framework and the Butcher framework of an algebra $B$ together
with a linear map $a:B\to B$ (cf.\cite{Bu}) is that we now
assume that the abelian algebra $B$ is in fact a Hopf algebra
while the map $a$ satisfies our Hochschild equation
\[
\Delta a(P)=a(P)\otimes 1+(id\otimes a)\Delta(P).
\]
The simplest example of such data is already given by the Picard
method of \cite{Bu} where the Hopf algebra structure (not provided
in \cite{Bu}) is given by
\[
\Delta x=x\otimes 1+1\otimes x,
\]
\[
a(P)(x)=\int_0^x P(u)du,
\]
cf.\cite{CK}. The difference between such data and a Runge-Kutta
method is that we now have translation invariance available.
Exactly as the Runge-Kutta method was producing an element in the
Butcher group, the above data give a homomorphism from the
underlying group (by the Milnor-Moore theorem) to the Butcher
group, a situation which is dual to our theorem on the
universality of ${\cal H}_R$.

We can now reformulate the ODE in general as an equation for $y\in
B\otimes E$,
\[
y=1\otimes\eta_0+(L\otimes id)f(y),
\]
where $\eta_0\in E$ is the initial datum  and $f(y)$ is uniquely
defined by  $f(y)(x)=f(y(x))$ for any $x\in spec(B)$.

This is very suggestive: somehow the usual solution curve for
an ODE should not be considered as an ultimate solution and 
the universal problem should be thought of as a refinement of the
idea of a scalar time parameter. We regard Butcher's work on the
classification of numerical integration methods as an impressive
example that concrete problem-oriented work can lead to far
reaching conceptual results.
\subsection{Iterated integrals and numbers from primitive
diagrams} The circle of ideas described so far certainly allows us
to come to terms with the combinatorics of the subtraction procedure
so that we can now concentrate only on those diagrams without
subdivergences, i.e.~the decorations at vertices of rooted trees.

The proper definition for the integral in the new calculus used in
Noncommutative Geometry is the Dixmier trace, i.e. the invariant
coefficient of the logarithmic divergence of an operator trace. A
superficial understanding of QFT would lead one to consider it as
far too limited a tool to confront the divergences of QFT. This
misgiving is based on the ignorance of the plain fact that the
divergence of a sub\-di\-ver\-gence-free diagram is a mere overall
logarithmic divergence, and that such diagrams, for which
Fig.(\ref{F2}) gives archetypical examples, appear at all
loop orders (the expert will notice that the decorations which are
to be used are actually Feynman graphs with appropriate
polynomial insertions which are the remainders when we shrink
subgraphs to points). \bookfig{}{!}{F2}{Diagrams free of
subdivergences and hence delivering a first order pole in
dimensional regularization. They are typically reflecting the
skeleton expansion of the theory.}{2.3}

In di\-men\-sio\-nal re\-gu\-la\-ri\-za\-tion the
de\-co\-ra\-tions appearing at the roo\-ted trees deliver a
first-order pole in $D-4$. In other words, the Dixmier trace, or
rather its residue guise $Res_-$, is perfectly sufficient to
disentangle and determine a general Feynman graph $G$, making full
use of the decomposition dictated by the Hopf algebra structure,
which gives the joblist \cite{BK} for the practitioner of QFT: the
list of diagrams which correspond to analytic expressions
suffering from merely an overall divergence.

Analyzing the corresponding residues of the decorations and their
number-theoretic properties  is a far reaching subject (discussed
in part by M.~Kontsevich in this issue) which by itself reveals
some interesting Hopf algebra structures, based on the algebraic
properties of iterated integrals.

Iterated integrals came to prominence with the work of K.T.~Chen \cite{chen}.
Essentially, they are governed by two rules,
Chen's lemma
\[
F_{r,t}^I=F_{r,s}^I+F_{s,t}^I+\sum_{I=(I^\prime,I^{\prime\prime})}
F_{r,s}^{I^\prime}F_{s,t}^{I^{\prime\prime}}
\]
and the shuffle product
\[
F_{r,s}^{I^\prime}F_{r,s}^{I^{\prime\prime}}=\sum_\sigma
F_{r,s}^{\sigma[I]},
\]
where the sum is over all $(p,q)$ shuffles of the symmetric group
$S_{p+q}$ acting on a string $I$ of, say, $n=p+q$ integers
parametrizing the one-forms which give the iterated integral as
their  integral over the $n$-dimensional standard simplex.

If we define $t^I$ to be a tree without sidebranches decorated
with these one-forms in the appropriate order, then each iterated
integral corresponds to a map from such a decorated tree to a real
number, so that Chen's lemma expresses the familiar group law of
the Butcher group in this special case. It is then the shuffle
product which guarantees that iterating one-forms in accordance
with a general rooted tree \cite{dchen} will not produce anything
new: any such integral is a linear combination of the standard
iterated integrals.

Most interestingly, the calculus of full perturbative QFT can be
understood as a calculus of generalized iterated integrals, where
the boundaries $r,s$ are to be replaced by elements of the Butcher
group: scalar functions of (decorated) rooted trees \cite{dchen}.
The group law  still holds naturally for such generalized
integrals, but the shuffle product only holds for the leading term
in the asymptotic expansion of bare diagrams \cite{DelKr}.

We hope that an investigation of this situation has far reaching
consequences for the understanding of the number-theoretic content
of Feynman diagrams, whose richness is underwritten by a wealth of
empirical data which provide a plethora of interesting numbers
like multiple zeta values \cite{zagier}, their alternating cousins
\cite{BK15}, and even sums involving non-trivial roots of unity in
numerators \cite{David1}.

It is gratifying then that the regularization of iterated
integrals based on forms $dx/x$, $dx/(1-x)$, providing the
iterated integrals representation of multiple zeta values, is in
full accord with the renormalization picture developed here. The
appearance of such numbers in the solution of the
Knizhnik-Zamolodchikov equation, and the relation to the Drinfel'd
associator upon renormalizing this solution so that it extends to
the unit interval, clearly motivates one to investigate the
differential equations which encapsulate the iterative structure
of Green functions in perturbative QFT, the bare and renormalized
Schwinger-Dyson equations.

Before we close this survey with a couple of remarks concerning
these equations, let us mention yet another application where the
freedom of having tree-indexed scales, hence elements of the
Butcher group, will prove essential.

The formulation of operator product expansion clearly takes place
in a configuration space of the same nature as the Fulton-MacPherson 
compactification \cite{FMPh}, known to be stratified by
rooted trees. However, there is a very important nuance
which can already be fully appreciated in the case of two points.
In that case, the Fulton-MacPherson space is simply the blow-up of
the diagonal in the space $X\times X$, whereas the geometric data
which encodes most of the semiclassical deformation aspect in a
diffeomorphism invariant manner are provided by
a smooth groupoid, called the
tangent groupoid in \cite{alain3}.

Essentially, the relation between the latter and the former is the
same as the relation between a linear space and the corresponding
projective space. That the transition to a linear space is a
crucial improvement can be appreciated from the fact that it is
only in the linear space that Fourier transform takes its full
power.

It is thus of great interest to extend the construction of the
tangent groupoid from the two-point case to the full set-up of the
configurations of  $n$ points. This clearly involves the freedom
of having scales available for any strata in the compactification,
hence again elements of the Butcher group. 
This is in full accord with the
use of tree-indexed parameters in the momentum space description
of the operator product expansion undertaken in \cite{dchen},
which clearly shows the necessity of allowing for the full linear
space to be able to describe the variety of renormalization
prescriptions employed by the practitioners of QFT.

\subsection{The Schwinger-Dyson Equations}
Recalling that equation (\ref{E1}) demands an understanding of the
full dress\-ed line element (cf.~Fig.( \ref{dress})) , we finally
consider Schwinger-Dyson equations. The propagator, the vertex and
the kernel functions provide a system of such   Schwinger-Dyson
equations, whose solution is at the heart of any Lagrangian QFT
\cite{BjD}. Though the fermion propagator, hence the line element,
comes to prominence in these equations due to gauge covariance, a
full solution is not yet known for any QFT of interest.

In the simplest form, the Schwinger-Dyson equation is typically
an equation of the form
\[ \Gamma(q)=\gamma+\int d^4k \Gamma(k)P(k)^2K(k,q),
\]
which is the structure of the Schwinger-Dyson
equation for the QED vertex at zero momentum transfer. Here, $P$
is a propagator, hence a geometric series in a self-energy (so
that the resulting equations are highly non-linear)  and $K$ is a
QED four-point kernel function, actually one which is a typical
generator (in its undressed form) of decorations upon closure of
two of its legs.

Nevertheless it is not difficult to identify an operator $L_K$
such that the solution is the formal series
\[
\Gamma(q)=\gamma+\sum_i L_K^i(\gamma),
\]
where the operator $L_K$ amounts to the operator $B_+$ combined
with all decorations provided by the kernel $K$. If the above
Schwinger-Dyson equation is for the bare vertex, the one for the
renormalized vertex ${\bf\Gamma}$ is obtained by multiplying it with
the counterterm $Z$ which amounts, in full accordance with the
$g_o^{-1}g$ notation used before, to the equation
\[
{\bf \Gamma}(q)=Z\gamma+\int d^4k {\bf \Gamma}(k)P(k)^2K(k,q),
\]
which neatly summarizes  the Hopf algebra structure of
perturbative QFT. From the above structure,
we conclude that Runge-Kutta methods are fully available
for this system.

Let us close this paper by noting an amusing coincidence: If we
restrict the kernel $K$ to the first two terms $K^{[1]},K^{[2]}$
in its skeleton expansion, restrict $P$ to the free propagators
and label each of the two terms $K^{[1]},K^{[2]}$ by noncommuting
variables $k_1$  and $k_2$ say, then we are naturally led to the
equation
\[
\Gamma(q)= \gamma+k_1\left[\int d^4k
\frac{1}{k^4}\Gamma(k)K^{[1]}\right] +k_2\left[\int d^4k
\frac{1}{k^4}\Gamma(k)K^{[2]}\right],
\]
which obviously involves a sum over all words in $k_1,k_2$ in its
solution and is fascinatingly close to the K-Z equation in two
variables considered before. \section*{Acknowledgements}
D.K.~thanks the IHES for extended hospitality and marvelous
working conditions. D.K.~also thanks the DFG for support as a
Heisenberg fellow.


\begin{thebibliography}{99}
\bibitem{BPHZ}N.N.~Bogoliubov, D.V.~Shirkov,
{\em Introduction to the theory of quantized fields}, 3rd ed.,
Wiley 1980;\\ K.~Hepp, Comm.Math.Phys.{\bf 2} (1966) 301-326;\\
W.~Zimmermann, Comm.Math.Phys.{\bf 15} (1969) 208-234.
\bibitem{BjD}
J.D.~Bjorken, S.D.~Drell, {\em Relativistic Quantum Fields},
McGraw Hill 1965.
\bibitem{David1} D.J.~Broadhurst, {\em Massive 3-loop
Feynman Diagrams reducible to SC${}^\star$ primitives of algebras
at the sixth root of unity}, to appear in Eur.Phys.J.C,
hep-th/9803091.
\bibitem{BK15} D.J.~Broadhurst, D.~Kreimer, Phys.Lett.B393 (1997) 403-412,
hep-th/9609128.
\bibitem{BK} D.J.~Broadhurst, D.~Kreimer, {\em
Renormalization automated by Hopf algebra}, to appear in
J.Symb.Comp., hep-th/9810087.
\bibitem{Br} C.~Brouder,
{\em Runge Kutta methods and renormalization},  hep-th/9904014.
\bibitem{Bu} J.C.~Butcher,
Math.Comp.26 (1972) 79-106.
\bibitem{Cayley} A.~Cayley {\em On the theory of the analytical forms 
called trees},
Phil.Mag.XIII (1857) 172-176.
\bibitem{chen} K.T.~Chen, Bull.Amer.Soc.83 (1977) 831-879.
\bibitem{alain1} A.~Connes, Comm.Math.Phys.182 (1996) 155-176, hep-th/9603053.
\bibitem{alain3} A.~Connes, {\em Noncommutative Geometry},
Acad.Press 1994.
\bibitem{alain2} A.~Connes, M.~Douglas, A.Schwarz,
JHEP 9802:003 (1998), hep-th/9711162.
\bibitem{CK} A.~Connes, D.~Kreimer, Comm.Math.Phys.199 (1998) 203-242,
hep-th/9808042.
\bibitem{CM} A.~Connes, H.~Moscovici, Comm.Math.Phys.198 (1998) 199-246,
math.DG/9806109.
\bibitem{EG} H.~Epstein, V.~Glaser, Ann.Inst.H.Poincar\'e 19 (1973) 211-295.
\bibitem{FMPh} W.~Fulton, R.~MacPherson, Annals of Math.(2).130 (1994) 183-225.
\bibitem{sasha} A.~Goncharov, {\em Polylogarithms in arithmetic and geometry},
Proc.~of ICM-94 (Z\"urich), Vol.1,2 374-387, Birkh\"auser 1995.
\bibitem{dhopf} D.~Kreimer, 
Adv.Theor.Math.Phys.2 (1998) 303-334, q-alg/9707029.
\bibitem{doverl} D.~Kreimer, {\em On Overlapping Divergences},
to appear in Comm.Math.Phys., hep-th/9810022.
\bibitem{dchen} D.~Kreimer, {\em Chen's Iterated Integral
represents the Operator Product Expansion}, hep-th/9901099.
\bibitem{DelKr} D.~Kreimer, R.~Delbourgo,
{\em Using the Hopf Algebra of QFT in calculations}, hep-th/9903249.
\bibitem{W} R.~Wulkenhaar, {\em On the Connes-Moscovici
Hopf algebra associated to the diffeomorphisms of a manifold},
preprint CPT Luminy April 1999.
\bibitem{zagier} D.~Zagier, {\em Values of zeta functions and their 
applications,} First
European Congress of Mathematics, Vol. II, 497-512,
Birkhauser, Boston,
1994.
\end{thebibliography}
\end{document}